\documentstyle[prl,aps,preprint,epsfig]{revtex}

\def\btabl{\begin{table}}   \def\etabl{\end{table}}
\def\bnn{\begin{eqnarray*}}   \def\enn{\end{eqnarray*}}
\def\btabu{\begin{tabular}}   \def\etabu{\end{tabular}}
\def\bec{\begin{displaymath}} \def\eec{\end{displaymath}}
\def\eqref#1{(\ref{#1})}

\newcommand{\bfig}{\begin{center}\begin{picture}}
\newcommand{\efig}[1]{\end{picture}\\{\small #1}\end{center}}

\newcommand{\mrm}[1]{\mbox{\rm #1}}
\newcommand{\beq}{\begin{equation}}
\newcommand{\eeq}{\end{equation}}
\newcommand{\nn}{\nonumber}
\newcommand{\bea}{\begin{eqnarray}}
\newcommand{\eea}{\end{eqnarray}}

\newcommand{\Eq}[1]{Eq.~(\ref{#1})}

\newcommand{\ea}{{ et al.}}

\newcommand{\np}[1]{{ Nucl. Phys. }{\bf #1}}
\newcommand{\plet}[1]{{ Phys. Lett. }{\bf #1}}
\newcommand{\pr}[1]{{ Phys. Rev. }{\bf #1}}
\newcommand{\prlet}[1]{{ Phys. Rev. Lett. }{\bf #1}}
\newcommand{\zp}[1]{{ Z. Phys. }{\bf #1}}

\newcommand{\ptp}[1]{{ Prog. Theor. Phys. }{\bf #1}} 
\newcommand{\arnps}[1]{{ Ann. Rev. Nucl. Part. Sci. }{\bf #1}}

\def\lsim{\mathrel{\vcenter{\hbox{$<$}\nointerlineskip\hbox{$\sim$}}}}
\def\gsim{\mathrel{\vcenter{\hbox{$>$}\nointerlineskip\hbox{$\sim$}}}}

\begin{document}

\preprint{\vbox{\baselineskip=13pt
\rightline{CERN-TH/2002-182}
\vspace*{2cm}
}}

\draft

\title{CP asymmetry in $B\to \phi K_S $ decays in 
left-right models  \\ 
and its implications on $B_s$ decays}

\author{\large M. Raidal}

\address{\vspace{0.5cm} 
{\small 
Theory Division, CERN, CH-1211, Geneva 23, Switzerland,\\
National Institute of Chemical Physics and Biophysics, 
Tallinn 10143 , Estonia }\vspace{0.5cm}
}


\maketitle 

\begin{abstract}
In left-right models the gluonic penguin contribution to $b\to s \bar s s$
transition is enhanced by $m_t/m_b$ due to the presence of (V+A) 
currents and by the larger values of loop functions than in the Standard 
Model. Together those may completely overcome the suppression due to small 
left-right mixing angle $\xi\lsim 0.013.$
Two independent new phases in the $B\to \phi K_S $ decay amplitude
appearing in a large class of left-right models may therefore modify the 
time dependent CP asymmetry in this decay mode by ${\cal O}(1)$ 
and explain the recent BaBar and Belle CP asymmetry measurements
in this channel. This new physics scenario implies observable 
deviations from the Standard Model also in $B_s$ decays which could be 
measured at upcoming Tevatron and LHC.
\end{abstract}

\vspace{2.5cm}

\leftline{CERN-TH/2002-182}

\leftline{August 2002}

\newpage 

The measurements of time-dependent asymmetries in $B\to J/\psi K_S$
have revealed CP violation in the $B$-system. The observed world 
average of $\sin 2\beta$ \cite{nir},
\bea 
\sin 2\beta_{J/\psi K}=0.734\pm 0.054,
\label{beta}
\eea
agrees well with the Standard Model (SM) prediction and
indicates that the Kobayashi-Maskawa (KM) 
mechanism \cite{km} is likely the dominant source of CP violation
also in this process. Nevertheless, this result does not exclude 
interesting CP violating new physics (NP) effects in other $B$ decays.
Since the decay $B\to J/\psi K_S$ $(b\to c\bar c s)$ is a tree level 
process in the SM, the NP contributions to its amplitude are naturally  
suppressed. However, at loop level NP may give large contributions 
to the $B^0$-$\bar B^0$  mixing as well as to the loop-induced 
decay amplitudes.  The former effects are universal to all $B^0$ 
decay modes and therefore constrained 
to be less than 20\% compared with the SM contribution \cite{nir}.
On the other hand, the effects of new physics in the
decay amplitudes are non-universal and can show up in the comparison of
the CP asymmetries in different decay modes \cite{gross}.

One of the most promising processes for NP searches widely considered 
in literature \cite{gross,susy,prl,susy2} 
is $B\to \phi K_S.$ In the SM the decay 
$b\to s \bar s s$ is one-loop effect and, according to the KM mechanism, 
the CP asymmetry in $B\to \phi K_S$ decay measures with high accuracy
the same quantity as  $B\to J/\psi K_S,$ namely $\sin 2\beta.$ 
The uncertainty for those processes in the SM is estimated to be 
\cite{gross,giw}
\bea
|\phi(B\to J/\psi K_S) - \phi(B\to \phi K_S)| \lsim {\cal O}(\lambda^2)\,,
\label{phi}
\eea  
where $\phi$ is the measured CP angle and $\lambda\approx 0.2.$  
Surprisingly, both BaBar \cite{babar} and Belle \cite{belle} 
obtain negative value for the CP asymmetry in this decay mode.
Their average result is 
\bea
\sin 2 \phi_{\phi K_S}=-0.39\pm 0.41 ,
\label{exp}
\eea 
where $\phi_{\phi K_S}\equiv \phi(B\to \phi K_S) $
denotes the measured CP angle. Despite of large statistical errors
those measurements establish a 2.7 $\sigma$ deviation from the
SM prediction $\sin 2\phi_{\phi K_S}=\sin 2\beta_{J/\psi K}$ 
and may indicate an effect of 
new physics. Since the deviation of \Eq{exp} from \Eq{beta} is very large, 
first analyses \cite{nir,hiller,datta,ciu} 
of this experimental result imply that one needs strongly enhanced gluonic
penguin contributions to the decay amplitude as in the generic 
supersymmetric models, non-standard flavour changing $Z$-boson couplings,
supersymmetry without R-parity etc. to account for such a large
deviation.

In this short note we would like to clarify that the result  \Eq{exp}
can actually be explained in a wide class of rather ordinary models 
from the flavour point of view: 
by the left-right symmetric models (LRSM) based on the
gauge group $\mrm{SU}(2)_R\times \mrm{SU}(2)_L\times \mrm{U}(1)_{B-L}$ 
\cite{lr}. Those models predict the existence of new charged gauge boson
$W_2$ with a mass  $M_2\gsim 1.6$ TeV \cite{beall} which may mix with
the SM gauge boson $W_1$ by the mixing angle $\xi\lsim 0.013$ \cite{lang}.
Due to the imposed discrete left-right symmetry the left- and right-handed 
Cabibbo-Kobayashi-Maskawa (CKM) matrices $V_L$ and $V_R,$ respectively, 
are related as $|V_L|=|V_R|$. However, their phases may differ from each 
other as happens in the models with spontaneous CP breaking 
\cite{cpk,cpb,bgnr,matias} which has six phases in $V_R.$
While the NP contribution to the $B^0$-$\bar B^0$ mixing is suppressed by
the heavy scale $M_2$ in this model, the gluonic penguin contributions to 
the  flavour changing decay $b\to s\bar s s$, which are proportional to
the mixing angle $\xi$, are enhanced by a large factor $m_t/m_b$ 
due to the presence of (V+A) interactions in the loop, and by another 
factor of four due to the larger values of Inami-Lim type loop functions.
Together those enhancement factors may overcome the suppression by 
$\xi,$ and the CP asymmetries in $B\to J/\psi K_S$ and $B\to K_S\phi$ 
may  differ from each other by order unity due to
the additional two independent phases in the $B\to K_S\phi$ decay 
amplitude. This scenario has important
consequences on the CP asymmetries in $b\to s\bar s s$ dominated $B_s$ decays 
such as $B_s\to \phi\phi$ which are predicted to be vanishing in the SM.
In the LRSM the BaBar and Belle result \Eq{exp} implies also the measurable 
CP asymmetries in $B_s$ decays at Tevatron and LHC.

Before going to the physics analyses let us comment on two relevant 
experimental and theoretical issues. Firstly, despite of the consistent
measurements of $\sin 2 \phi_{\phi K_S}$ by two experiments (with the
direct CP asymmetry consistent with zero \cite{belle} as reported by Belle),
the time dependent CP asymmetry in $B\to \eta' K_S$ decay which also has 
a $s\bar s$ component is consistent with $\sin 2\beta_{J/\psi K}$ 
in \Eq{beta}. 
Since $B\to \eta' K_S $ is not as clean process as $B\to \phi K_S$,
this result can still be tolerated together with the NP in $B\to \phi K_S$
\cite{hiller}. Furthermore, as argued in Ref. \cite{ciu}, due to the 
depencence on the final state hadronic matrix elements the relation between
CP asymmetries in $B\to \eta' K_S$ and $B\to \phi K_S$ may be non-trivial.
Nevertheless, improving 
experimental accuracy is going to impose serious constraints on
the idea of NP in decay amplitudes if this inconsistency persists. 
Secondly, the results in this paper are valid in LRSM with relatively low 
$SU(2)_R$ breaking scale which could be motivated also by neutrino physics
\cite{low}.
It is shown in Ref.\cite{bgnr} that spontaneous CP violation in the LRSM
with minimal Higgs sector leads to the SM only with fine tunings while, in 
general, there exist additional light Higgs multiplets. With low $M_2$
the fine tunings are not severe and light Higgs multiplets may give additional
NP contributions to the CP asymmetries. The low-scale scenario has also
a potential to be tested directly at lepton \cite{lep} 
and hadron \cite{lhc} colliders.
In addition, one can always extend
the model to non-minimal one or, as the simplest possibility, just abandon 
the imposed-by-hand discrete left-right symmetry which results in
unconstrained $V_R$ \cite{lang}. 
We do not study those model building issues here.
Instead, we address generic phenomenological consequences 
of the model which are consistent with all the present experimental bounds.

CP violation in $B^0$ decays  
takes place  due to the interference between mixing and decay. 
The corresponding CP asymmetry depends on the parameter $\lambda$
defined as \cite{quinn}
\bea
\lambda=\left( \sqrt{\frac{M_{12}^*-\frac{i}{2}\Gamma_{12}^*}
{M_{12}-\frac{i}{2}\Gamma_{12}}}\right)\frac{\bar A}{A}=
\frac{q}{p}\frac{\bar A}{A}\,,
\label{lambda}
\eea
where $A$ and $\bar A$ are the amplitudes of $B^0$ and $\bar B^0$
decay to a common CP eigenstate, respectively. With a good accuracy
$|q/p|=1$ and the $B$-$\bar B$ mixing phase is given by 
$q/p=e^{-2i\phi_M}.$ Neglecting the direct CP asymmetry 
one has $|\lambda|=1$ and  $\bar A/A=e^{-2i\phi_D} $ 
gives the phase in the decay amplitude.  In this case the time dependent 
CP asymmetry takes a particularly simple form
\bea
 a_{CP}(t)=-Im\lambda\sin(\Delta Mt)=\sin 2(\phi_M+\phi_D)\sin(\Delta Mt), 
\label{acp}
\eea
where $\Delta M$ is the mass difference between the two physical states.
From \Eq{lambda} and \Eq{acp} it is clear that any new physics effect 
in the mixing will translate into  $\phi_M\to \phi_M+\delta_M$ and will be
 universal to 
all decays while the effect in the decay, $\phi_D\to\phi_D+\delta_D,$
will depend on the decay mode. As the NP in $\phi_M$ is already constrained 
to be below 20\%, we proceed with studying the NP in decay amplitudes and 
comment on $\phi_M$ effects later.

The charged current Lagrangian in the LRSM  is given by 
\bea
{\cal L}_{cc} &=& \frac{g}{\sqrt{2}} \;\overline{u} \left( \cos\xi V_L 
\gamma^\mu P_L - e^{i \omega} \sin\xi V_R \gamma^\mu P_R \right) d \;
W_{1\mu} + \nn \\
&&
\frac{g}{\sqrt{2}}\; \overline{u} \left( 
e^{-i\omega} \sin\xi V_L 
\gamma^\mu P_L + \cos\xi V_R \gamma^\mu P_R \right) d \;
W_{2\mu}  +  \mbox{H.c.},
\eea
where $P_{L,R} \equiv (1 \mp \gamma_5)/2,$ 
$W_1,$ $W_2$ are the charged vector boson fields with the masses
$M_1,$ $M_2,$ respectively, $\xi$ denotes their mixing and $\omega$
is a CP phase.

The flavour changing decay $b\to s\bar ss$ is induced by 
the  QCD-, electroweak-
and magnetic penguins.  The dominant contribution comes from the QCD penguins
with top quark in the loop.
It is also known \cite{fleish1} that the electroweak penguins 
decrease  the decay rate by about 30\%.
We shall add all those contribution to the QCD improved effective Hamiltonian. 
We start with
the effective Hamiltonian due to the gluon exchange
describing the decay $b\to s\bar ss$
at the scale $M_1$  
\bea
H^0_{eff}= -\frac{G_F}{\sqrt{2}}\frac{\alpha_s}{\pi} V_L^{ts*}V_L^{tb}
\left( \bar s\left[\Gamma_\mu^{LL}+\Gamma_\mu^{LR} \right] T^a b\right)
\left( \bar s \gamma^\mu T^a s \right),
\label{heffmw}
\eea
where
\bea
\Gamma_\mu^{LL}&=&E_0(x)\gamma_\mu P_L +
2i\frac{m_b}{q^2}E_0'(x)\sigma_{\mu\nu}q^\nu P_R, \nn \\
\Gamma_\mu^{LR}&=&2i\frac{m_b}{q^2}\tilde E_0'(x)
[A^{tb} \sigma_{\mu\nu}q^\nu P_R +
A^{ts*} \sigma_{\mu\nu}q^\nu P_L ],
\eea
the $\Gamma_\mu^{LR}$ term describes the new dominant
left-right contribution due to the mixing angle $\xi,$
 and 
\bea
A^{tb}=\xi \frac{m_t}{m_b} \frac{V_R^{tb}}{V_L^{tb}}e^{i\omega}\equiv
\xi \frac{m_t}{m_b} e^{i\sigma_1}, \;\;\;\;\;
A^{ts}=\xi \frac{m_t}{m_b} \frac{V_R^{ts}}{V_L^{ts}} e^{i\omega}\equiv
\xi \frac{m_t}{m_b} e^{i\sigma_2}.
\label{a}
\eea
Note that the phases $\sigma_{1,2}$
are independent and can take any value in the range $(0,2\pi).$
 The functions $E_0(x),$ $E_0'(x)$ and
$\tilde E_0'(x)$ are Inami-Lim type functions \cite{lim} of 
$x=m_t^2/M_1^2$ and are given by
\bea 
E_0(x)&=&-\frac{2}{3}\ln x + \frac{x(18-11x-x^2)}{(12(1-x)^3)}+
\frac{x^2(15-16 x+4x^2)}{(6(1-x)^4)}\ln x, \nn \\
E_0'(x)&=& \frac{x(2+5x-x^2)}{(8(x-1)^3)}-
\frac{3x^2}{(4(x-1)^4)}\ln x, \nn \\
\tilde E_0'(x)&=& -\frac{(4+x+x^2)}{(4(x-1)^2)}+
\frac{3x}{(2(x-1)^3)}\ln x . 
\eea
Notice that $\tilde E_0'(x_t)$ is numerically about factor of four larger 
than the SM function $E_0'(x_t).$
Together with the $m_t/m_b$ enhancement in $A^{tb}$, $A^{ts}$
this practically overcomes the left-right suppression by small $\xi$
and allows large CP effects in the decay amplitude due to the 
new phases $\sigma_{1,2}$. We note that the analogous effect is also 
responsible for the enhancement of gluonic penguins in general 
supersymmetric models \cite{susy2}.

To calculate $B$ meson decay rates at the energy scale $\mu=m_b$ 
in the leading logarithm (LL) 
approximation we adopt the procedure from Ref. \cite{misiak}.
Using the operator product expansion to integrate out the heavy 
fields, and to calculate  the LL Wilson coefficients $C_i(\mu)$  
we run them with the renormalization 
group equations from the scale of $\mu=W_1$ 
down to the scale $\mu=m_b$ (since the contributions of $W_2$
are negligible we  start immediately from the $W_1$ scale).  
Because the new physics appears only in the  
magnetic dipole  operators we can safely take over some well-known
results from the SM studies. 
Therefore the the LRSM effective Hamiltonian  should 
include only these new terms which mix with the gluon and 
photon dipole operators under QCD renormalization. 
We work with the effective Hamiltonian 
\bea
H_{eff}&=&\frac{G_F}{\sqrt{2}} \left[
V_L^{us*}V_L^{ub}
\sum_{i=1,2} C_i(\mu) O^u_i(\mu) +
V_L^{cs*}V_L^{cb}
\sum_{i=1,2} C_i(\mu) O^c_i(\mu) \right. \nn \\
&-& \left. 
V_L^{ts*}V_L^{tb}\left(
\sum_{i=3}^{12} C_i(\mu) O_i(\mu) +
 C^{\gamma}_7(\mu) O^{\gamma}_7(\mu) + C^{G}_7(\mu) O^{G}_7(\mu)
\right)\right] + (C_iO_i \to C'_iO'_i)\,,
\eea
where $O_{1,2}$ are the standard current-current operators,
$O_3$-$O_{6}$ and $O_7$-$O_{10}$ are the standard QCD and EW penguin
operators, respectively, and  $O_7^\gamma$ and $O_8^G$ are the
standard photonic and gluonic magnetic operators, respectively.
They can be found in the literature (e.g. Ref. \cite{ag,cct}) and 
we do not present them here.
The new operators to be added, $O_{11,12},$ 
are analogous to the current-current
operators  $O_{1,2}$ but with different chiral structure  \cite{misiak}
\bea
O_{11}&=& \frac{m_b}{m_c}(\bar s_\alpha\gamma^{\mu}(1-\gamma_5) c_\beta )
(\bar c_\beta \gamma_\mu (1+\gamma_5)b_\alpha), \nn \\
O_{12}&=& \frac{m_b}{m_c}(\bar s_\alpha\gamma^{\mu}(1-\gamma_5)c_\alpha )
(\bar c_\beta \gamma_\mu (1+\gamma_5)b_\beta).
\eea
 Due to the left-right symmetry the operator basis is doubled by
including operators  $O'_i$ which can be obtained from $O_i$ 
by the replacements $P_L\leftrightarrow P_R.$

Because the new physics affects only the Wilson coefficients
$C_7^\gamma,$ $C_8^G$ and $C_7^\gamma,$ $C_8^G$ it is sufficient 
to consider the basis $O_{1-6},O^\gamma_7,O^G_8,O_{11,12}$ + $(O\to O')$
for calculating them in the LL precision. The relevant matching conditions
can be found in \cite{matias} and we do not present them here.
The $20\times 20$ anomalous dimension  matrix decomposes into two identical
$10\times 10$ sub-matrices. The SM $8\times 8$ 
sub-matrix of the latter one can be found in Ref. \cite{rome} and
the rest of the entries have been calculated  in Ref. \cite{misiak}.
In the LL approximation the low energy Wilson 
coefficients for five flavours are given by 
\bea
C_i(\mu=m_b)=\sum_{k,l}(S^{-1})_{ik}(\eta^{3\lambda_k/46})S_{kl} C_l(M_1),
\eea
where the $\lambda_k$'s in the exponent of $\eta=\alpha_s(M_1)/\alpha_s(m_b)$
are the eigenvalues of the anomalous dimension matrix over $g^2/16\pi^2$
and the matrix $S$ contains the corresponding eigenvectors. 
The result for the gluonic magnetic coefficients relevant for our studies
is \cite{prl}
\bea
C^G_8(m_b)&=&\eta^{\frac{14}{23}}(E_0'(x)+A^{tb}\tilde E_0'(x)) +
\sum_{i=1}^5 h'_i \eta^{p'_i}\,, \\
C'^G_8(m_b)&=&\eta^{\frac{14}{23}}A^{ts*}\tilde E_0'(x) \,,
\eea
where
$h'_{i}=(0.8623,$ -0.9135, 0.0209, 0.0873, -0.0571) and
$p'_{i}=(14/23,$ 0.4086, 0.1456, -0.4230, -0.8994).
Using $\Lambda^{(5)}_{\bar MS}=225$ MeV and $\mu=\bar m_b(m_b)=4.4$ GeV
we find numerically 
$C^\gamma_7=-0.331-0.523 A^{tb}$, $C'^\gamma_7=-0.523 A^{ts*},$
$C^G_8=-0.156-0.231 A^{tb}$ and $C'^G_8=-0.231 A^{ts*}.$

To calculate the hadronic matrix element  
$\langle O \rangle\equiv\langle K_S\phi|O|B \rangle$
for the  $B\to K_S\phi$ decay amplitude we use the factorization 
approximation which has been extensively discussed in the
literature \cite{bsw,ag,cct} and we do not repeat it here.
However, treating  the most relevant matrix element for our studies, 
\bea
 \langle O^G_8\rangle &=&-\frac{2\alpha_s}{\pi}\frac{m_b}{q^2}
\langle (\bar s_\alpha i\sigma_{\mu\nu}q^\mu P_R T^a_{\alpha\beta} b_\beta )
(\bar s_\gamma \gamma^\nu T^a_{\gamma\delta} s_\delta)\rangle \,,
\label{o8}
\eea
where $q^\mu$ is the momentum transfered by the gluon to the $(\bar s,s)$ 
pair, is non-trivial. Following \cite{ag} the result is \cite{matias}
\bea
 \langle O^G_8\rangle &=&-\frac{\alpha_s}{4\pi}
\frac{m_b}{\sqrt{\langle q^2\rangle}}
\left[\langle O_4 \rangle+\langle O_6 \rangle - \frac{1}{N_c}
\left( \langle O_3 \rangle+\langle O_5 \rangle\right) \right] \,,
\eea
and similarly for $ \langle O'^G_8\rangle .$ 
The parameter $\langle q^2\rangle$ introduces certain uncertainty
into the calculation. In the literature its value is varied in the range
$1/4 \lsim \langle q^2\rangle/m_b^2 \lsim 1/2$ \cite{hou}. 

In the factorization approach
the amplitude $A\equiv\langle H_{eff}\rangle$ of the decay  $B\to \phi K_S$
takes a form \cite{cct}
\bea
A(B\to \phi K_S)=-\frac{G_F}{\sqrt{2}}V_L^{tb}V_L^{ts*}2 \left[
a_3+a_4+a_5-\frac{1}{2}(a_7+a_9+a_{10})\right]
X^{(B K,\phi)}\,,
\label{A}
\eea 
where $X^{(B K,\phi)}$ stands for the factorizable hadronic
matrix element which exact form is irrelevant for us 
since it cancels out in CP asymmetries.
The coefficients $a_i$ are given by
\bea
a_{2i-1}= C^{eff}_{2i-1}+\frac{1}{N_c}C^{eff}_{2i}\,,
\qquad\qquad
a_{2i}= C^{eff}_{2i}+\frac{1}{N_c}C^{eff}_{2i-1}\,,
\eea
where the QCD improved coefficients $C^{eff}_i$ can be found in \cite{matias}.
Using $\sqrt{\langle q^2\rangle}=m_b/\sqrt{2},$ $\xi=0.01$ and 
$m_t/m_b=60$ we obtain for the LL QCD improved amplitude 
\bea
A (B\to \phi K_S) =-\frac{G_F}{\sqrt{2}}V_L^{tb}V_L^{ts*}
2\left[-0.016+0.0035 \left(e^{i\sigma_1}+e^{-i\sigma_2}\right)\right]
X^{(BK,\phi)} \,.
\label{ares}
\eea 
The maximum effect occurs for phases $\sigma_1=-\sigma_2=\pi/2+\delta_D.$ 
Numerically we get $(\bar A/A)_{max}=e^{\pm 0.91 i}.$ We recall that 
this estimate is obtained for the most conservative $\langle q^2\rangle $.
Using more optimistic $\sqrt{\langle q^2\rangle}=m_b/2$ the NP effect
is increased to $(\bar A/A)_{max}=e^{\pm 1.3 i}.$ 
According to \Eq{lambda} and \Eq{acp} the NP phase in $\bar A/A$
can change the CP asymmetry by of order unity.
Therefore, consistency with the BaBar and Belle result \Eq{exp} can be 
obtained in the LRSM.

Our scenario is expected to influence also the decay $B\to X_s\gamma$
\cite{misiak,bsg,atwood} which is given by penguin diagrams too. 
However, $b\to s \gamma$ is induced only by right-projected operators.
It is  possible \cite{atwood} that due to the cancellation between 
the $LL$ and $LR$ contributions both the total rate $\Gamma$ and 
the CP asymmetry in this process  correspond to the SM predictions
while the NP CP effect in $B\to \phi K_S$ decay are still of order one.
The same conclusion holds also in supersymmetric models \cite{bsgsusy}.

Our explanation to the BaBar and Belle measurements of the time-dependent
CP asymmetry in  $B\to \phi K_S$ decays due to the NP in the decay amplitude
has important consequences for the $B_s$ decays do be measured at
Tevatron and LHC. One of the cleanest process is the pure penguin induced 
decay $B_s\to\phi\phi$ ($b\to s\bar s s$).
Its branching ratio is large, of the order 
$B(B_s\to\phi\phi)\sim {\cal O}(10^{-5})$ \cite{cct}
and the pollution from other SM
diagrams is estimated to be of order ${\cal O}(1)$\% \cite{gross}. 
Since the CP asymmetries in this mode should vanish in the SM,  the decay 
$B_s\to\phi\phi$ should provide very sensitive tests of the SM
at hadron machines. 

Formally the $B_s\to\phi\phi$ amplitude is also given by \Eq{A}
but with a proper hadronic matrix element $X^{(B_s\phi,\phi)}.$
However, in the factorization approximation the hadronic
matrix elements of the operators $O_i$ and $O'_i$ depend on the spin of
the decay products. For $B_s\to PP,\;VV$ where $P$ and $V$ denote
any pseudo-scalar and vector meson, respectively, one has 
 $\langle O_i\rangle=-\langle O'_i\rangle$ while for the decays of the 
type $B_s\to PV$ one has $\langle O_i\rangle=\langle O'_i\rangle.$
Therefore the magnetic penguin contributions which give NP effects 
may have different signs in different processes.
Using the same numerical input as before we obtain
\bea
A(B_s\to\phi\phi)=-\frac{G_F}{\sqrt{2}}V_L^{tb}V_L^{ts*}
2 \left[-0.016+0.0035\left(e^{i\sigma_1}-e^{-i\sigma_2}\right)\right]
X^{(B_s\phi,\phi)} \,.
\label{a1}
\eea 
Unless $\sigma_1=-\sigma_2$, ${\cal O}(1)$  deviation from the SM prediction 
$a^{SM}_{CP}(B_s\to\phi\phi)=0$ can be expected with the maximal 
results $(\bar A/A)_{max}=e^{\pm 0.91 i (\pm 1.3 i)}$ as before.
Should $\sigma_1=-\sigma_2$ indeed be the case, one has to search for
the CP asymmetries in the processes $B_s^0\to \eta\rho^0$, $B_s^0\to \pi\phi$ 
which are $PV$ type but may have large tree level contributions.
However, this is unjustified fine tuning and NP effects of order ${\cal O}(1)$ 
can be expected both in $B\to K_S\phi$ and $B_s\to\phi\phi.$
Therefore Tevatron or LHC should be able to test our scenario 
also in $B_s$ decays.

Let us finally comment on the NP contribution to the $B$-$\bar B$ mixing
in LRSM. The mixing phase $\phi_M$ can be modified as 
$\phi_M=\phi_M^{SM}+ \delta_M$ where \cite{cpb}
\bea
\delta_M=\arctan\left(\frac{\kappa\sin\sigma}
{1+\kappa\cos\sigma}\right) \,,
\label{deltam}
\eea 
and $e^{i\sigma} \simeq  - (V_{R,td} V_{R,tb}^*)/(V_{L,td} V_{L,tb}^*).$
Keeping only the $W_{1,2}$ contributions,
the LL QCD improved result for $\kappa=|M_{12}^{LR}|/|M_{12}^{LL}|$
is \cite{cpb}  $\kappa = F(M_2) \left(1.6 \mbox{TeV}/M_2 \right)^2,$ 
where $F(1.6\,\mrm{TeV})=0.2.$
Therefore the NP contribution to the mixing phase may be non-negligible
but is estimated to be below 20\% of the SM value.

In conclusion, if the measured discrepancy between the time-dependent
CP asymmetries in $B\to J/\psi K_S$ and $B\to \phi K_S$ decays
is due to new physics, it can be explained, consistently with all 
experimental bounds,  by the enhanced gluonic penguin contribution to 
the $B\to \phi K_S$ decay amplitude in the LRSM.
This scenario implies also large CP asymmetry in the decay $B_s\to\phi\phi$
(and also in $B_s^0\to \eta\rho^0$, $B_s^0\to \pi\phi$)
which can be tested in upcoming Tevatron and LHC.

\vspace{1.5cm}

\begin{center}
{\bf Acknowledgements}
\end{center}

I am very grateful to G. Barenboim for earlier collaboration on this 
topic and to G. D'Ambrosio for discussions.
This work is partially supported by EU TMR contract No. 
HPMF-CT-2000-00460 and by ESF grant No. 5135.

\end{document}